\documentclass[prd,reprint,showkeys]{revtex4}%
\usepackage{textcomp}
\usepackage{amssymb}
\usepackage{eurosym}
\usepackage{amsmath}
\usepackage{epsfig}
\usepackage{graphics}
\usepackage{color}
\usepackage{graphicx}
\usepackage[font={footnotesize,it}]{caption}
\usepackage[colorlinks=true,
            linkcolor=red,
            urlcolor=blue,
            citecolor=green]{hyperref}
\usepackage{pdfpages}
\usepackage{pmboxdraw}
\usepackage{calrsfs}
\usepackage{amsfonts}%
\providecommand{\U}[1]{\protect\rule{.1in}{.1in}}
\makeatletter
\@ifundefined{textcolor}{}
{
\definecolor{BLACK}{gray}{0}
 \definecolor{WHITE}{gray}{1}
 \definecolor{RED}{rgb}{1,0,0}
 \definecolor{GREEN}{rgb}{0,1,0}
 \definecolor{BLUE}{rgb}{0,0,1}
 \definecolor{CYAN}{cmyk}{1,0,0,0}
 \definecolor{MAGENTA}{cmyk}{0,1,0,0}
 \definecolor{YELLOW}{cmyk}{0,0,1,0}
}

\makeatother
\begin{document}
\title{Spectroscopy of $z=0$ Lifshitz Black Hole}
\author{G. Tokgoz}
\email{gulnihal.tokgoz@emu.edu.tr}
\author{I. Sakalli}
\email{izzet.sakalli@emu.edu.tr}
\date{\today }


\begin{abstract}
We studied the thermodynamics and spectroscopy of a $4$-dimensional, $z=0$
Lifshitz black hole ($Z0$LBH). Using the Wald's entropy formula and the
Hawking temperature, we derived the quasilocal mass of the $Z0$LBH. Based on the
exact solution to the near-horizon Zerilli equation of the massive scalar
waves, we computed the quasinormal modes of the $Z0$LBH via employing the adiabatic invariant quantity for the $Z0$LBH. This study shows that the entropy and area spectra of the $Z0$LBH are equally spaced.
\end{abstract}
\affiliation{Physics Department , Eastern Mediterranean University, Famagusta, Northern
Cyprus, Mersin 10, Turkey}
\maketitle

\section{Introduction}
Ever since the publication of the seminal papers of Bekenstein and Hawking \cite{main1,main2,Beks1}, a black hole (BH) has been known to have a quantum origin. It was discussed in detail by Bekenstein \cite{Beks2,Beks3,Beks4,Beks5} that entropy of a BH ($S_{BH}$) should be quantized in discrete levels. The proportionality between entropy and BH area ($\mathcal{A}_{BH}$) is justified from the adiabatic invariance \cite{Efst} properties of area, that is, $S_{BH}=\frac{\mathcal{A}_{BH}}{4}$. Therefore, $\mathcal{A}_{BH}$  should also be quantized in equidistant levels to account for the discrete $S_{BH}$. Bekenstein's heuristic quantization arguments \cite{Beks2,Beks3} showed that for the family of Schwarzschild BHs, $\mathcal{A}_{BH}$ should have the following discrete, equidistant spectrum:

\begin{equation}
\mathcal{A}_{BH}=8\pi n\hbar,\text{ \ \ \ \ \ \ }n=0,1,2,... \label{1}%
\end{equation}

where $\epsilon$ is known as the undetermined dimensionless constant.
According to Eq. (1), the minimum increase in the horizon area becomes
$\Delta\mathcal{A}_{\min}=8\pi\hbar$ \cite{myMag,Vagenas,Medved}. Following
the seminal works of Bekenstein, new methods have been developed to derive
the entropy and the area spectra of the numerous BHs (see \cite{myGua} and references
therein). Among them, Maggiore's method \cite{myMag} fully supports Bekenstein's result (1). In fact, \cite{myMag} was based on Kunstatter's
study \cite{myKuns} in which the adiabatic invariant quantity ($I_{adb}$) is expressed as follows:

\begin{equation}
I_{adb}=\int\frac{dM}{\Delta\omega}, \label{2}%
\end{equation}

where $\Delta\omega=\omega_{n-1}-\omega_{n}$ denotes the transition frequency
between the subsequent levels of a BH having mass $M$. Further, Eq. (2) was
generalized to the hairy BHs (massive, charged, and rotating ones) as follows
(see \cite{myKwon} and references therein):%

\begin{equation}
I_{adb}=\int\frac{\mathcal{T_H}dS_{BH}}{\Delta\omega}, \label{3}%
\end{equation}

where $\mathcal{T_H}$ is the temperature of the BH. Bohr--Sommerfeld quantization rule \cite{BSR} states that $I_{adb}$ acts as a quantized quantity ($(I_{adb}\simeq n\hbar)$) when the highly excited modes ($n\rightarrow\infty$) are considered. In such a case, the imaginary part of the frequency dominates the real part of the frequency ($\omega_{I}\gg\omega_{R}$), implying that $\Delta\omega\simeq\Delta\omega_{I}$. Meanwhile, for the first time, Hod \cite{myHod1,myHod2} argued that QNMs \cite {Corless,Fernando} can be used for computing transition frequency. Hod's arguments inspired Maggiore, who considered the Schwarzschild BH as a highly damped harmonic oscillator (i.e., $\Delta\omega\simeq\Delta\omega_{I}$) and managed to rederive Bekenstein's original result (1) using a different method. Today, there are numerous studies in the literature in which the Maggiore's method (MM) has been employed for various BHs (see for instance
\cite{myLi,myKothawala,Ortega,Sakalli1,Sakalli2,myAnn}).

This study mainly explores the entropy and area spectra of a four-dimensional Lifshitz BH \cite{Lu} possessing the particular dynamical exponent $z=0$. To analyze the physical features of the $Z0$LBH geometry, we first calculate its quasilocal mass $M_{QL}$ \cite{myBY} and temperature via the Wald's entropy \cite{Eune} and statistical Hawking temperature formula \cite{Wald1,Wald0}, respectively. QNM calculations of the $Z0$LBH must be performed to successfully implement the MM. To this end, we consider the Klein–Gordon equation (KGE) for a massless scalar field in the $Z0$LBH background. Separation of the angular and the radial equations yields a Schrödinger type wave equation or the so-called Zerilli equation \cite{Chandrasekhar}. Plots of the potential show that the effective potential may diverge beyond the BH horizon in some cases; thus, in the far region, the QNMs might not be perceived by the observer. Hence, considering the possibility of not all QNMs being detected by the distant observer, we assume that there is a probe located in the vicinity of the BH. The probe receives the frequencies of the QNMs and transmits them to the distant observer. In the framework of this scenario, we focus our analysis on the near-horizon (NH) region and use the boundary condition, which causes the outgoing waves to vanish at the event horizon. After getting the NH form of the Zerilli equation, we show that the radial equation is reduced to a confluent hypergeometric (CH) differential equation \cite{Abramowitz}. Via performing some manipulations on the NH solution and using the pole structure of the Gamma functions \cite{Abramowitz}, we show how one finds out the QNMs as in \cite{Sakalli3,Sakalli4,Setare1,Setare2,Cuadros}. The imaginary part of the QNMs is used in Eq. (3), and the quantum spectra of entropy and area are obtained.

Following statements elaborate on the organization of this study. In Sec. II, we briefly introduce the $Z0$LBH metric. In addition, we present the derivation of $M_{QL}$  of the $Z0$LBH based on the Wald's entropy formula. Section III is devoted to the separation of the KGE and finding the effective potential $V_{eff}\left(r\right)$. Next, we solve the NH Zerilli equation and show how QNMs are calculated. Then, we compute the entropy and area spectra of the $Z0$LBH. Finally, we draw our conclusions in Sec. IV. (Throughout this work, the geometrized unit system is used: $G=c=k_{B}=1$ and $\ell_{\rho}^{2}=\hbar$.)

\section{$Z0$LBH Spacetime}

In this section, we introduce the four-dimensional Lifshitz spacetimes \cite{Lu} and its special case, that is, $Z0$LBH. The conformal gravity (CG) covers the gravity theories, which are invariant under Weyl transformations. CG, which is adopted  to static and asymptotically Lifshitz BH solutions, has received intensive attraction from the researchers studying condensed matter and quantum field theories \cite{Kachru}.
The Lifshitz BHs are invariant under anisotropic scale and characterize the gravitational dual of strange metals \cite{Hartnoll}. Lifshitz BHs are asymptotically described by the following line-element \cite{Lu}

\begin{equation}
ds^{2}=-\frac{r^{2z}}{l^{2z}}dt^{2}+\frac{l^{2}}{r^{2}}dr^{2}+\frac{r^{2}%
}{l^{2}}d\vec{x}^{2}, \label{4}%
\end{equation}

where $l$, $z,$ and $\vec{x}$ denote the length scale in the geometry, the
dynamical exponent, and the $2D$ spatial vector, respectively. The action of
the Einstein--Weyl gravity \cite{Lu} is given by%

\begin{equation}
\mathcal{S}=\frac{1}{2\widetilde{\kappa}^{2}}\int\sqrt{-g}d^{4}x(R-2\Lambda+\frac{1}%
{2}\alpha\left\vert Weyl\right\vert ^{2}), \label{5}%
\end{equation}

where $\widetilde{\kappa}^{2}=8\pi G$ , $\left\vert Weyl\right\vert
^{2}=R^{\mu\nu\rho\sigma}R_{\mu\nu\rho\sigma}-2R^{\mu\nu}R_{\mu\nu}+\frac
{1}{3}R^{2}$, and $\alpha=\frac{z^{2}+2z+3}{4z(z-4)}$ (constant), which
diverges ($\alpha=\infty)$ with $z=0$ and/or $z=4$.

It is shown that the static asymptotically Lifshitz BH solutions exist in the CG theory for both $z=4$ and $z=0$; however, when $z=3$ and $z=4$, Lifshitz BHs appear in the Horava--Lifshitz gravity \cite{Lu,Catalan}. Now, we focus on the $Z0$LBH of the CG theory whose metric is given by

\begin{equation}
ds^{2}=-fdt^{2}+\frac{4dr^{2}}{r^{2}f}+r^{2}d\Omega_{2,k}^{2}, \label{6}%
\end{equation}

where the metric function $f$ is defined by%

\begin{equation}
f=1+\frac{c}{r^{2}}+\frac{c^{2}-k^{2}}{3r^{4}}. \label{7}%
\end{equation}

In the above metric, $d\Omega_{2,k}^{2}=\frac{dx^{2}}{1-kx^{2}}+(1-kx^{2}%
)dy^{2}$. Herein, $k=0$, $k=1$, and $k=-1$ stand for $2$-torus, $2$-sphere (which
can be expressed as $d\theta^{2}+\sin^{2}\theta d\phi^{2}$), and unit
hyperbolic plane, respectively. Metric (6) conformally describes an AdS BH
when $k=-1$, however $k=1$ corresponds to a dS BH. The metric solution has a
curvature singularity at $r=0$, which becomes naked for $k=0$. There is an event horizon for $k=\pm1$ solution expressed as follows:

\begin{equation}
r_{h}^{2}=\frac{1}{6}\left(  \sqrt{3(4-c^{2})}-3c\right)  . \label{8}%
\end{equation}

Note that the requirement of $r_{h}^{2}\geq0$ is conditional on this
inequality: $-2\leq c<1$. Throughout this study, we consider the choice of
$k=1$ and $c=-1$. Thus, the metric function $f$ becomes%

\begin{equation}
f=1-\frac{r_{h}^{2}}{r^{2}}. \label{9}%
\end{equation}

Thus, at spatial infinity, the Ricci and Kretschmann scalars of the $Z0$LBH
can be found as follows%

\begin{align}
R  &  =R_{\lambda}^{\lambda}\sim\frac{5(c^{4}-2c^{2}+1)}{12r^{8}},\nonumber\\
K  &  =R^{\mu\nu\rho\sigma}R_{\mu\nu\rho\sigma}\sim\frac{25(c^{4}-2c^{2}%
+1)}{12r^{8}}. \label{10}%
\end{align}

By performing the surface gravity calculation \cite{main1,main2,Wald0}, we obtain%

\begin{equation}
\kappa=\frac{1}{2}. \label{11}%
\end{equation}

Therefore, the Hawking or BH temperature \cite{Wald0} of the
$Z0$LBH reads%

\begin{equation}
\mathcal{T_H}=\frac{\kappa}{2\pi}=\frac{1}{4\pi}. \label{12}%
\end{equation}

\subsection{Mass Computation of $Z0$LBH via Wald's Entropy Formula}

The GR unifies space, time and gravitation and the gravitational force is
represented by the curvature of the spacetime. Energy conservation is a sine
qua non in GR as well. Because the metric (6) of $Z0$LBH represents a
non-asymptotically flat geometry, one should consider the quasilocal mass
$M_{QL}$ \cite{myBY}, which measures the density of matter/energy of the
spacetime. In this section, we shall employ Wald's entropy calculation
\cite{Wald1,Wald0} and derive the $M_{QL}$ using Wald's entropy formula. To
this end, we follow the study of Eune and Kim \cite{Eune}.

Starting with the timelike Killing vector $\xi^{\mu},$ which describes the
symmetry of time translation in a spacetime, the Wald's entropy is expressed
by \cite{Wald1,Wald0} as%

\begin{equation}
S_{BH} =\frac{2\pi}{\kappa}%
{\displaystyle\int\limits_{\Sigma}}
d^{2}x\sqrt{h}\beta, \label{13}%
\end{equation}

where%

\begin{align}
\beta &  =\epsilon_{\mu\nu}J^{\mu\nu},\nonumber\\
\epsilon_{\mu\nu}  &  =\frac{1}{2}\left(  n_{\mu}u_{\nu}-n_{\nu}u_{\mu
}\right)  . \label{14}%
\end{align}

Here, $\kappa$ and $h$ are the surface gravity and the induced metric on a
hypersurface $\Sigma$ of the horizon ($2D$-sphere with $\sqrt{h}%
=r^{2}sin\theta$), respectively. $u_{%
\mu
}$ is the four-vector velocity defined as the proper velocity of a fiducial
observer moving along the orbit of $\xi^{\mu}=\gamma\frac{\partial}{\partial
t}$ (where $\gamma$ is a normalization constant) which must satisfy $g_{\mu
\nu}\xi^{\mu}\xi^{\nu}=-1$ at spatial infinity. Thus, one can immediately see
that $\gamma=1.$ $J^{\mu\nu}$ is called the Noether potential
\cite{Lopes,Parikh}, which is given by%

\begin{equation}
J^{\mu\nu}=-2\Theta^{\mu\nu\rho\sigma}\left(  \nabla_{\rho}\xi_{\sigma
}\right)  +4\left(  \nabla_{\rho}\Theta^{\mu\nu\rho\sigma}\right)  \xi
_{\sigma}, \label{15}%
\end{equation}

with%

\begin{equation}
\Theta^{\mu\nu\rho\sigma}=\frac{1}{32}(g^{\mu\rho}g^{\nu\sigma}-g^{\mu\sigma
}g^{\nu\rho}). \label{16}%
\end{equation}

The surface gravity $\kappa$ can be calculated by \cite{Wald0}%

\begin{equation}
\kappa=\lim_{r\rightarrow r_{h}}\sqrt{\frac{\xi^{\mu}\nabla_{\mu}\xi_{\nu}%
\xi^{\rho}\nabla_{\rho}\xi^{\nu}}{-\xi^{2}}}=\frac{1}{2}. \label{17}%
\end{equation}

To have an outward unit normal vector $n_{\mu}$ on $\Sigma$, the equality
$n_{\mu}n^{\mu}=1$ must also be satisfied. Hence, one can get%

\begin{equation}
n_{r}=\frac{1}{n^{r}}=\frac{2}{\sqrt{(r^{2}-r_{h}^{2})}}. \label{18}%
\end{equation}

On the contrary, $u_{\mu}$ is the four-vector velocity and is given by%

\begin{equation}
u^{\mu}=\frac{1}{\alpha}\xi^{\mu}, \label{19}%
\end{equation}

with%

\begin{equation}
\alpha=\sqrt{-\xi^{\mu}\xi_{\mu}}. \label{20}%
\end{equation}

Therefore, the non-zero components of $\epsilon_{\mu\nu}$\ are found to be%

\begin{equation}
\epsilon_{tr}=-\epsilon_{rt}=-\frac{1}{r}. \label{21}%
\end{equation}

In sequel, the four-vector velocity reads%

\begin{equation}
u^{t}=\frac{r}{\sqrt{r^{2}-r_{h}^{2}}}. \label{22}%
\end{equation}

The non-zero components of $\Theta^{\mu\nu\rho\sigma}$ (16) are obtained as follows%

\begin{align}
\Theta^{trtr}  &  =\Theta^{rtrt}=-\Theta^{trrt}=-\Theta^{rttr}=\frac{-r^{2}%
}{128\pi},\nonumber\\
\Theta^{\theta r\theta r}  &  =\Theta^{r\theta r\theta}=-\Theta^{\theta
rr\theta}=-\Theta^{r\theta\theta r}=\frac{r^{2}-r_{h}^{2}}{128\pi r^{2}%
},\nonumber\\
\Theta^{\phi r\phi r}  &  =\Theta^{r\phi r\phi}=-\Theta^{\phi rr\phi}%
=-\Theta^{r\phi\phi r}=\frac{r^{2}-r_{h}^{2}}{128\pi r^{2}\sin^{2}\theta
},\nonumber\\
\Theta^{\theta\phi\theta\phi}  &  =\Theta^{\phi\theta\phi\theta}%
=-\Theta^{\theta\phi\phi\theta}=-\Theta^{\phi\theta\theta\phi}=\frac{1}{32\pi
r^{4}\sin^{2}\theta},\nonumber\\
\Theta^{t\phi t\phi}  &  =\Theta^{\phi t\phi t}=-\Theta^{t\phi\phi t}%
=-\Theta^{\phi tt\phi}=\frac{-1}{32\pi\left(  r^{2}-r_{h}^{2}\right)  \sin
^{2}\theta},\nonumber\\
\Theta^{\theta t\theta t}  &  =\Theta^{t\theta t\theta}=-\Theta^{\theta
tt\theta}=-\Theta^{t\theta\theta t}=\frac{-1}{32\pi\left(  r^{2}-r_{h}%
^{2}\right)  }. \label{23}%
\end{align}

One can verify that $\nabla_{\rho}\Theta^{\mu\nu\rho\sigma}=0$. The latter
result yields that the second term of the Noether potential vanishes. Therefore,
we have%

\begin{equation}
J^{\mu\nu}=-2\Theta^{\mu\nu\rho\sigma}\left(  \nabla_{\rho}\xi_{\sigma
}\right)  =-2\Theta^{\mu\nu\rho\sigma}C_{\rho\sigma}, \label{24}%
\end{equation}

The non-zero components of $C_{\rho\sigma}=\nabla_{\rho}\xi_{\sigma}$ are as follows%

\begin{equation}
C_{tr}=-C_{rt}=-\gamma r_{h}^{2}/r^{3}. \label{25}%
\end{equation}

After substituting those findings into Eq. (15), we find the nonzero
components of the Noether potential:%

\begin{equation}
J^{tr}=-J^{rt}=\dfrac{-r_{h}^{2}}{32\pi}. \label{26}%
\end{equation}

Thus, from Eq. (16), $\beta$\ is found as%

\begin{equation}
\beta=\frac{r_{h}^{2}}{16r^{2}\pi}, \label{27}%
\end{equation}

and in sequel computing the entropy through the integral formulation (13), we obtain%

\begin{equation}
\mathcal{S}=\pi r_{h}^{2}=\frac{\mathcal{A}_{h}}{4}. \label{28}%
\end{equation}

The above result is fully consistent with the Bekenstein-Hawking entropy. The
quasilocal mass $M_{QL}$ can be derived from this entropy by integrating the
first law of thermodynamics $dM_{QL}=\mathcal{T_H}dS_{BH}$. After some manipulation,
one easily finds the following result%

\begin{equation}
M_{QL}=\frac{r_{h}^{2}}{4}. \label{29}%
\end{equation}

Consequently, the metric function (9) of the metric reads%

\begin{equation}
f=1-\frac{4M}{r^{2}}. \label{30}%
\end{equation}

\section{QNMs and Spectroscopy of $Z0$LBH}

In this section, we shall study the QNMs and the entropy of a perturbed
$Z0$LBH via MM \cite{myMag}. QNMs of a considered BH can be derived by
solving the eigenvalue problem of the KGE with the proper boundary conditions.
The boundary condition at the horizon implies that there are no outgoing waves
at the event horizon (i.e., only ingoing waves carry the QNMs at the event
horizon) and the boundary condition at spatial infinity imposes that only
outgoing waves are allowed to survive at spatial infinity; therefore, those
waves would be detected by an asymptotic observer. On the contrary, in this section, we construct a Gedanken experiment in which it is assumed that there exists a probe in the vicinity of the horizon and that probe is able to detect and analyze the ingoing waves and then transmit their QNM frequencies to a distant observer. Thus, we will show how one can have the QNMs of the $Z0$LBH using their NH boundary condition. For this purpose, we first consider the massive KGE:%

\begin{equation}
\frac{1}{\sqrt{-g}}\partial_{\mu}\left(  \sqrt{-g}g^{\mu\nu}\partial_{\nu
}\right)  \Psi-m^{2}\Psi=0, \label{31}%
\end{equation}

where $\Psi$ adopts the ansatz for the above wave equation chosen as%

\begin{equation}
\Psi=\frac{1}{r}F(r)e^{i\omega t}Y_{lm}(\theta,\phi), \label{32}%
\end{equation}

in which $F(r)$ is the function of $r$ and $Y_{lm}(\theta,\phi)$\ represents the
spherical harmonics with the eigenvalue $-l(l+1)$. After performing some
straightforward calculations, the radial part of the KGE reduces to a
Schrödinger-like equation or the so-called Zerilli equation
\cite{Chandrasekhar}:%

\begin{equation}
\left[  -\frac{d}{dr^{\ast2}}+V_{eff}\right]  F(r)=\omega^{2}F(r), \label{33}%
\end{equation}

where $V_{eff}$ and $r^{\ast}$ are called the effective potential and the
tortoise coordinate, respectively. The tortoise coordinate $r^{\ast}$ can be
found by the following integral%

\begin{equation}
r^{\ast}=2\int\frac{dr}{rf}, \label{34}%
\end{equation}

which results in%

\begin{equation}
r^{\ast}=\ln\left(  \frac{r^{2}}{r_{h}^{2}}-1\right)  . \label{35}%
\end{equation}

One may check that the limits of $r^{\ast}$ admit the following:%

\begin{align}
\lim_{r\rightarrow r_{h}}r^{\ast}  &  =-\infty,\nonumber\\
\lim_{r\rightarrow\infty}r^{\ast}  &  =\infty. \label{36}%
\end{align}

The effective potential seen in the Zerilli equation (33) is obtained as%

\begin{equation}
V_{eff}=f\left\{  \frac{l(l+1)}{r^{2}}+\frac{1}{4}\left[  f+r\frac{df}%
{dr}\right]  +m^{2}\right\}  , \label{37}%
\end{equation}

which admits these limits:%

\begin{align}
\lim_{r\rightarrow r_{h}}V_{eff}(r)  &  =0,\nonumber\\
\lim_{r\rightarrow\infty}V_{eff}(r)  &  =m^{2}+\frac{1}{4}. \label{38}%
\end{align}

Clearly, the potential never terminates at the spatial infinity, instead, it
tends to diverge when the scalar particle is very massive, that is, $m\rightarrow
\infty$.%

\begin{figure}[h]
\includegraphics[width=0.57\textwidth]{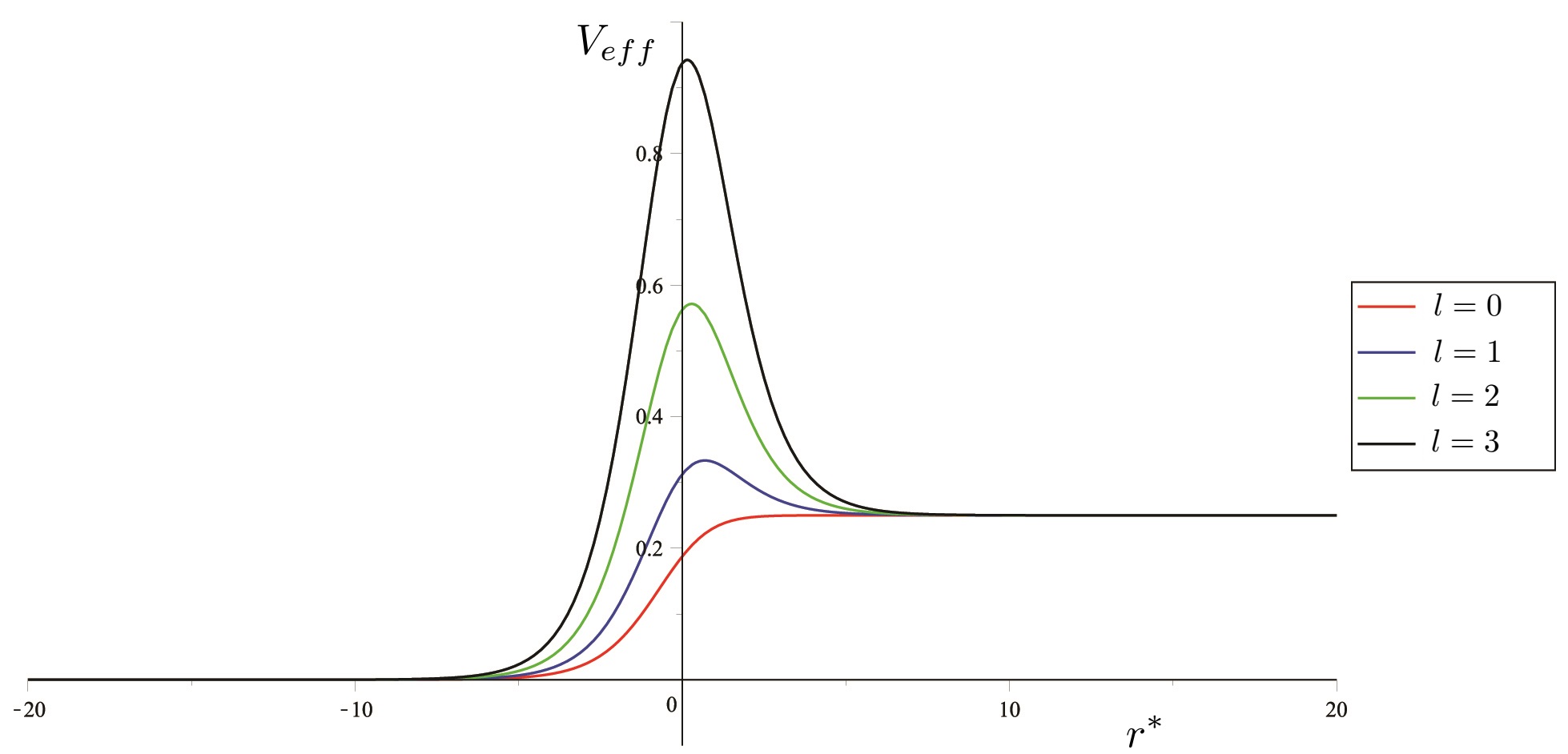}
\caption{{\ Effective potential versus tortoise coordinate graph for various orbital quantum numbers.}}
\end{figure}

\subsection{Entropy/Area Spectra of $Z0$LBH}

In this section, we shall nudge (perturb) the $Z0$LBH by the massive scalar
fields propagating near the event horizon and read their corresponding
QNM frequencies.

We can expand the $Z0$LBH's metric function $f$ to a series around $r_{h}$ and
express it in terms of the surface gravity $\kappa$ as%

\begin{equation}
f=f(r_{h})+f^{\prime}(r_{h})(r-r_{h})+O[(r-r_{h})^{2}]\approx2\kappa y,
\label{39}%
\end{equation}

where $y=r-r_{h}$ and prime ($^{\prime}$) denotes the derivative with respect
to $r$. After substituting Eq. (39) into Eq. (37) and performing Taylor
expansion, the NH form of the Zerilli potential is obtained as%

\begin{equation}
V=4\kappa Gy\left[  G^{2}\left(  1-2Gy\right)  l\left(  l+1\right)
+\kappa(1+2Gy)+m^{2}\right]  , \label{40}%
\end{equation}

with the parameter $G=\frac{1}{r_{h}}$. The tortoise coordinate in the NH region becomes $r^{\ast}\simeq\frac{1}{2\kappa}\ln y,$ which enables us to find the NH form of the Zerilli equation:%

\begin{equation}
\left[  -4\kappa^{2}y\left(  y\frac{d^{2}}{dy^{2}}+\frac{d}{dy}\right)
+V-\omega^{2}\right]  F(y)=0. \label{41}%
\end{equation}

The solutions to the Zerilli equation (41) can be expressed in terms of the CH
functions of the first and second kinds \cite{Abramowitz} as follows%

\begin{equation}
F(y)=y^{\frac{i\omega}{2\kappa}}e^{-\frac{z}{2}}\left[  C_{1}M(a,b,z)+C_{2}%
U(a,b,z)\right]  , \label{42}%
\end{equation}

with the parameters%

\begin{align}
a  &  =\frac{\lambda}{\sqrt{\delta}}+\frac{b}{2},\nonumber\\
b  &  =1+i\frac{\omega}{\kappa},\nonumber\\
z  &  =2iG\sqrt{\delta}y, \label{43}%
\end{align}

where%

\begin{align}
\lambda &  =\frac{1}{2}\left\{  1+\frac{1}{\kappa}\left[  l\left(  l+1\right)
G^{2}+m^{2}\right]  \right\}  ,\nonumber\\
\delta &  =2\left[  l\left(  l+1\right)  \frac{G^{2}}{\kappa}-1\right]  .
\label{44}%
\end{align}

With the aid of the limiting forms of the CH functions \cite{Abramowitz}, one
can find the NH limit of the solution (42) as%

\begin{equation}
F(y)\sim\left[  C_{1}+C_{2}\frac{\Gamma(1-b)}{\Gamma(1+a-b)}\right]
y^{\frac{i\omega}{2\kappa}}+C_{2}\frac{\Gamma(b-1)}{\Gamma(a)}y^{-\frac
{i\omega}{2\kappa}}. \label{45}%
\end{equation}

We can alternatively represent Eq. (45) in terms of $r^{\ast}$ ($y\simeq
e^{2\kappa r^{\ast}}$). Thus, the NH ingoing and outgoing waves are distinguished:%

\begin{equation}
\Psi\sim\left[  C_{1}+C_{2}\frac{\Gamma(1-b)}{\Gamma(1+a-b)}\right]
e^{i\omega(t+r^{\ast})}+C_{2}\frac{\Gamma(b-1)}{\Gamma(a)}e^{i\omega
(t-r^{\ast})}. \label{46}%
\end{equation}

Imposing the boundary condition that the outgoing waves must vanish at the
horizon, the solution having coefficient $C_{2}$ should be terminated. By
using the pole structure of the Gamma function for the denominator of the
second term, the outgoing waves vanish for $a=-n$ ($n=0,1,2,..$). The latter
remark yields the frequencies of the QNMs of the $Z0$LBH as%

\begin{equation}
\omega_{n}=2\kappa\left[  (n+\frac{1}{2})i+\frac{\lambda}{\sqrt{\delta}%
}\right]  , \label{47}%
\end{equation}

where $n$ is known as the overtone quantum number \cite{Hod3}. Accordingly,
the transition frequency between two highly excited subsequent states
($\omega_{I}\gg\omega_{R}$) is easily obtained as%

\begin{equation}
\Delta\omega\approx\Delta\omega_{I}=2\kappa=\frac{4\pi\mathcal{T_H}}{\hbar}.
\label{48}%
\end{equation}

Subsequently, using the adiabatic invariant quantity (3) and Bohr--Sommerfeld quantization rule%

\begin{equation}
I_{adb}=\frac{\hbar}{4\pi}\int\frac{dM}{\mathcal{T_H}}=\frac{\hbar}{4\pi
}S_{BH}=\hbar n, \label{49}%
\end{equation}

we can read the entropy and area spectra of the $Z0$LBH as follows%

\begin{align}
S_{BHn}  &  =\frac{\mathcal{A}_{n}}{4\hbar}=4\pi n,\nonumber\\
\mathcal{A}_{n}  &  =16\pi\hbar n. \label{50}%
\end{align}

Therefore, the minimum spacing of the BH area becomes%

\begin{equation}
\Delta\mathcal{A}_{\min}=16\pi\hbar. \label{51}%
\end{equation}

Our finding is in agreement with the Bekenstein's conjecture \cite{Beks5}, and the equispacing of the entropy/area spectra of the $Z0$LBH support the Kothawala et al.'s hypothesis \cite{myKothawala}, which states that BHs should have equally spaced area spectrum in Einstein's gravity theory.

\section{Conclusions}
In this study, the quantum spectra of the $Z0$LBH were studied using the MM, which is based on the adiabatic invariant quantity (3). After separating the radial and angular parts of the massive KGE on the $Z0$LBH background, we have found the Zerilli equation and its corresponding effective potential (37). We have checked the behaviors of the potential around horizon and at the spatial infinity [see Eq. (38)]. In addition, we have depicted the effective potential for different values and have shown that the potential never terminates at the spatial infinity. The Zerilli equation associated to the $Z0$LBH is approximated to a CH differential equation. We have derived QNMs of the $Z0$LBH using the pole feature of the Gamma functions. The MM is applied for the highly excited modes and the entropy/area spectra are calculated. Both spectra are evenly spaced and independent of the BH parameters. On the contrary, we obtained a different area equispacing ($\epsilon=16\pi$)  than the usual value of the dimensionless constant ($\epsilon=8\pi$ \cite{myMag,Vagenas,Medved}). However, as shown by Hod \cite{myHod2}, the spacing between two adjacent levels might be different depending on which method is applied for studying the BH quantization. Besides, our findings are in agreement with both the Bekenstein's conjecture \cite{Beks5} as well as Wei et al. and Kothawala et al.'s studies \cite{Wei,myKothawala}.

In addition, $M_{QL}$ of the $Z0$LBH is also investigated via the Wald's entropy formula (28) by integrating the total energy (mass). The result obtained is in agreement with the BH thermodynamics \cite{BHterm}. Our next target is to study the Dirac QNMs of the 4-dimensional $Z0$LBH and analyze the spin effect on the area/entropy quantization.

\end{document}